\documentstyle[apjfonts,emulateapj]{article}
\slugcomment{To appear in ApJ, Jun 2001 issue}
\lefthead{KIM \& MORRIS}
\righthead{DIFFUSION OF INNER GALACTIC BULGE STARS}
\def\spose#1{\hbox to 0pt{#1\hss}}
\newcommand\lsim{\mathrel{\spose{\lower 3.0pt\hbox{$\mathchar"218$}}
     \raise 2.0pt\hbox{$\mathchar"13C$}}}
\newcommand\gsim{\mathrel{\spose{\lower 3.0pt\hbox{$\mathchar"218$}}
     \raise 2.0pt\hbox{$\mathchar"13E$}}}
\newcommand\msun{{\rm \,M_\odot}}

\begin{document}
\title{SPATIAL DIFFUSION OF STARS IN THE INNER GALACTIC BULGE}
\author{Sungsoo S. Kim \& Mark Morris}
\affil{Division of Astronomy \& Astrophysics, University of
California, Los Angeles, CA 90095-1562;\\
sskim@astro.ucla.edu, morris@astro.ucla.edu}

%%%%%%%%%%%%%%%%%%%%%%%%%%%%%%%%%%%%%%%%%%%%%%%%%%%%%%%%%%%%%%%%%%%%%%%%%%%%%%%%
\begin{abstract}
Star formation in the inner few hundred pc of the Galactic bulge occurs
in a flattened molecular layer called the central molecular zone (CMZ).
\markcite{SM96}Serabyn \& Morris (1996) suggest that the star
formation in the CMZ has been sustained for the lifetime of the Galaxy,
and that the resulting agglomeration of stars formed in the CMZ has resulted
in the prominent $r^{-2}$
stellar density cusp at the Galactic center having about the same physical
extent as the CMZ.  This ``central cusp'' is somewhat less flat than the CMZ;
thus the population of stars formed in the CMZ appears to have diffused out to
larger latitudes.  We hypothesize that such vertical diffusion is driven
by the scattering of stars off the giant molecular clouds (GMC) in the
CMZ, and perform numerical simulations of the scattering between stars
and GMCs in the presence of the non-axisymmetric background potential.
The simulation results show that the time scale for an initially flattened
stellar population to achieve an aspect ratio of the observed OH/IR stars
in the inner bulge, 1 to 2 Gyr, agrees well with the estimated age of those
OH/IR stars.
\end{abstract}


%%%%%%%%%%%%%%%%%%%%%%%%%%%%%%%%%%%%%%%%%%%%%%%%%%%%%%%%%%%%%%%%%%%%%%%%%%%%%%%%
\section{INTRODUCTION}
\label{sec:introduction}

It has long been known that the central bulge of the Galaxy has
a distinct stellar density profile which approximately follows an
$r^{-2}$ law (\markcite{BN68}Becklin \& Neugebauer 1968).  This $r^{-2}$
cusp has often been assumed to be merely the innermost part of the
elderly bulge (\markcite{EdZ94}Evans \& de Zeeuw 1994, among others).
However, the fact that the Central Molecular Zone (CMZ) has the same
maximum identifiable extent in Galactic longitude as this stellar cusp
($\sim 200$~pc) led \markcite{SM96}Serabyn \& Morris (1996) to propose that
star formation has been steadily occurring in the CMZ
throughout the lifetime of the Galaxy.

\begin{figure*}
%Fig 1
%\centerline{\epsfxsize=16cm\epsfbox{cmz2.ps}}
\plotfiddle{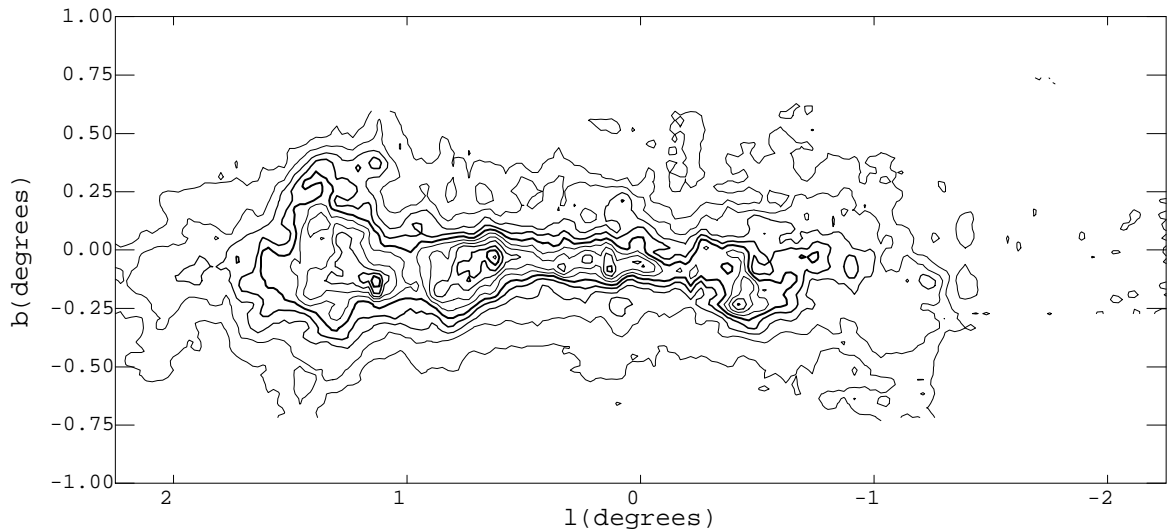}{7cm}{90}{63}{63}{232}{-90}
\caption
{\label{fig:cmz}
Integrated intensity of $^{12}$CO, $J=1$--0 emission, from the
AT\&T Bell Labs survey of the inner Galactic bulge region
(Uchida, Bally, \& Morris, unpublished work).  Contour lines were
drawn at intervals of $250 \, {\rm K \cdot km \, s^{-1}}$, starting from
$250 \, {\rm K \cdot km \, s^{-1}}$.
}
\end{figure*}

The CMZ is a massive reservoir of molecular clouds unparalleled
in the Galaxy.  It is an active region of star formation with
a number of very young stellar clusters and ample ionized gas
occupying the central $\sim 400 \,{\rm pc} \times 100 \, {\rm pc}$ region
($l \times b$; see Figure~\ref{fig:cmz}).
The clouds within the CMZ have relatively high density
($n \gsim 10^4 \,{\rm cm^{-3}}$), high
volume filling factor ($f \gsim 0.1$), and significantly elevated
temperatures (30-200~K, typically $\sim 70$~K; \markcite{He93}H\"uttemeister
et al. 1993, among others), and their total mass is estimated at $5-10 \times
10^7 \msun$ (\markcite{SM96}Serabyn \& Morris 1996),
representing roughly 10\% of our Galaxy's neutral gas
content.  The inevitable angular momentum loss in an orbiting gas disk
results in gas inflow along the Galactic plane from the outer bulge to the
CMZ.  \markcite{SM96}Serabyn \& Morris enumerate the mechanisms for
the angular momentum loss as follows: 1) shear viscosity in the differentially
rotating gas disk, 2) shocks associated with a bar potential, 3) cloud-cloud
collisions, 4) dynamical friction of giant molecular clouds by field
stars (\markcite{Se91}Stark et al. 1991), 5) magnetic field viscosity
(\markcite{M96}Morris 1996), and 6) dilution of specific angular momentum
by stellar mass loss material raining down out of the slowly rotating
Galactic bulge (\markcite{JB94}Jenkins \& Binney 1994).

Both the stellar population ($r^{-2}$ cusp) and the CMZ are considerably
flattened along the Galactic plane.  While the aspect ratio of the bulge is
1.6--1.7 (\markcite{KDF91}Kent, Dame, \& Fazio 1991; \markcite{We94}Weiland
et al. 1994), those of the stellar population and the
CMZ are $\sim 2.2$ (\markcite{CWG90}Catchpole, Whitelock, \& Glass 1990)
and $\sim 4$ (\markcite{SM96}Serabyn \& Morris 1996), respectively.
The mass of the stellar population within the $\sim 100$-pc extent of the
CMZ, $\sim 10^9 \, {\rm M_\odot}$,
falls near the lower end of the range predicted for the stellar mass emerging
from the CMZ over the Galaxy's lifetime, and Serabyn \& Morris consider
this as additional support for the association between the stellar
population and the CMZ.

Here we focus our attention on the difference of the aspect ratios between
the $r^{-2}$ stellar population and the CMZ.  In the context of
the sustained star formation hypothesis, one may interpret the
observation that the aspect ratio of the stellar population is
smaller than the CMZ in terms of the
vertical diffusion of stars formed in the flatter CMZ by gravitational
perturbations.

The vertical heating of the stars born in the Galactic disk due to
gravitational perturbations has been studied by many authors, especially
for heating by giant molecular clouds (GMCs).
\markcite{SS51}\markcite{SS53}Spitzer \& Schwarzschild (1951, 1953)
first proposed that random encounters of the stars formed in a thin disk
with massive interstellar clouds would heat up the stellar populations
(the presence of GMCs was not known until $\sim 20$~years later).
They used a Fokker-Planck model to calculate the evolution of the
velocity dispersions.  \markcite{L84}Lacey (1984) performed
similar calculations but with a consideration of vertical motions
of stars.  \markcite{V83}\markcite{V85}Villumsen (1983, 1985)
approached the same problem by integrating the equation of motion
for a tracer population of non-self-gravitating stars evolving
in the gravitational fields of the fixed background (disk and halo)
and of the GMCs.  On the other hand, \markcite{BW67}Barbanis \&
Woltjer (1967) proposed that transient spiral density waves could
heat up the stellar populations.  A series of papers
(\markcite{CS85}Carlberg \& Sellwood 1985; \markcite{C87}Carlberg 1987;
\markcite{SK91}Sellwood \& Kahn 1991)
has explored this mechanism numerically.  However, neither of the
above two mechanisms alone was able to explain the observations.  The models
involving scattering by the GMCs predicted the ratio of stellar velocity
dispersions perpendicular to the plane and towards the Galactic center,
$\sigma_z / \sigma_R$, to be larger than that observed.  They also
predicted that the velocity dispersion increases with time as
$\sim t^{0.25}$, which is substantially slower than implied by the
observations, $\sim t^{0.5}$ (\markcite{W77}Wielen 1977).  Meanwhile,
spiral structure was found to be highly inefficient in increasing
$\sigma_z$.  Thus, both mechanisms needed to be considered to account
for the observations, and the study of the stellar heating under the
combined influence of both GMCs and spiral structure was made by
\markcite{JB90}Jenkins \& Binney (1990), whose simulations were
found to agree with observations of spiral structure material near the Sun.

Contrary to the case of the Galactic disk, the kinematic evolution
of a population of stars formed in the CMZ via heating processes
has not been studied.  The kinematic environments in the inner bulge
differ from those in the disk mainly in three respects:
1) the background gravitational potential in the inner bulge is
neither axisymmetric nor plane-parallel; 2) the presence of density
waves strong enough to significantly heat the stellar population seems
unlikely (weakly damped modes may be present though; see, e.g.,
\markcite{W94}Weinberg 1994); and 3) the dynamical time scale
in the inner bulge is considerably shorter than in the disk.
Because of 1), we may not apply the analytic study of the kinematic
evolution of the disk stars directly to the inner bulge.
In the present paper, we hypothesize that the stellar population
formed in the CMZ diffuses mainly by scattering off the GMCs and we study
the kinematic and spatial evolution of stars in a flattened configuration
by following the trajectory of non-self-gravitating particles
under the gravitational potential of the background (inner bulge and
nucleus) and the GMCs, similarly to the work by \markcite{V85}Villumsen
(1985) for the case of the Galactic disk.  We choose to approach this
problem numerically, in order to obtain more tangible results.
%The approximations required to formulate a tractable analytic treatment
%would lead to solutions of dubious applicability.

If the scattering off the GMCs causes the stellar population to diffuse
out from the inner bulge faster than the rate at which stars are formed
in the CMZ, the resulting stellar population would have a
significantly larger extent than the CMZ, instead of forming a
configuration having a scale similar to the CMZ, such as the $r^{-2}$ stellar
population that we currently observe.
On the other hand, if the diffusion is highly inefficient, the resulting
stellar population from the sustained star formation in the CMZ would
have a configuration very similar to that of the CMZ.  Since the time scale
of the stellar diffusion in the inner bulge is an important parameter,
the calculation of the vertical evolution of the stellar population
and its comparison with observations are essential for assessing
our hypothesis.  We compare our simulation results with the distribution
of OH/IR stars, one of few object categories in the inner
bulge that currently provides approximate information on the age.

Our simulation models are described in \S~\ref{sec:models}.  We show and
discuss our results in \S\S~\ref{sec:results} and \ref{sec:discussion}.
Finally, our findings are summarized in \S~\ref{sec:summary}.

%%%%%%%%%%%%%%%%%%%%%%%%%%%%%%%%%%%%%%%%%%%%%%%%%%%%%%%%%%%%%%%%%%%%%%%%%%%%%%%%
\section{MODELS}
\label{sec:models}

As discussed in \S~1, the background potential in the inner bulge
is neither axisymmetric nor plane-parallel, which complicates any
analytic approach for the kinematic evolution of stars within it.
We therefore study the problem by directly integrating the orbital motion of
non-self-gravitating test particles.  The total integration time
of our simulations is either 1 or 3 Gyr.  Our model parameters discussed
below are given in Table~\ref{table:runs}.

\begin{deluxetable}{cccccccc}
\scriptsize
\tablecolumns{8}
\tablewidth{0pt}
\tablecaption{Model Parameters for Diffusion Simulations
\label{table:runs}}
\tablehead{
\colhead{} &
\colhead{$\Omega_p$} &
\colhead{$M_{GMC}$} &
\colhead{} &
\colhead{$\epsilon_{GMC}$} &
\colhead{} &
\colhead{$\Phi_0$} &
\colhead{} \\
\colhead{Model} &
\colhead{(${\rm km \, s^{-1} \, kpc^{-1}}$)} &
\colhead{($M_\odot$)} &
\colhead{$\alpha_{GMC}$} &
\colhead{(pc)} &
\colhead{$\alpha$} &
\colhead{(${\rm pc^2 \, G^{-1} \, Myr^{-2}}$)} &
\colhead{$z_0$}
}
\startdata
1& 65& $2.5\times 10^7$& -1& 10& -1.8& $3.1\times 10^4$& 0.8\\
2& 65& $2.5\times 10^7$& -1& 10& -1.8& $3.1\times 10^4$& {\bf 0.7}\\
3& 65& $2.5\times 10^7$& -1& 10& {\bf -1.5}& ${\bf 4.5\times 10^3}$& 0.8\\
4& 65& $2.5\times 10^7$& -1& 10& -1.8& ${\bf 7.8\times 10^4}$& 0.8\\
5& {\bf 50}& $2.5\times 10^7$& -1& 10& -1.8& $3.1\times 10^4$& 0.8\\
6& 65& ${\bf 5.0\times 10^7}$& -1& 10& -1.8& $3.1\times 10^4$& 0.8\\
7& 65& $2.5\times 10^7$& {\bf 0}& 10& -1.8& $3.1\times 10^4$& 0.8\\
8& 65& $2.5\times 10^7$& -1& {\bf 5}& -1.8& $3.1\times 10^4$& 0.8\\
\tablecomments{Model 1 is our standard model.  Variations from model~1 are
denoted by boldfaces.  For all models, $\sigma_{v,xy} = 20 \, {\rm km \,
s^{-1}}$, $\sigma_{v,z} = 25 \, {\rm km \, s^{-1}}$, $N_{GMC}=20$,
$y_0=0.9$, and $a_0 = r_0 = 1$~pc.}
\enddata
\normalsize
\end{deluxetable}

\subsection{GMCs}
\label{sec:gmcs}

Many groups have conducted radio observations of molecular clouds in
the inner bulge (see \markcite{M97}Morris 1997 for a list of such
observations).  Here, for the GMC parameters in our simulations,
we adopt a CO line survey by \markcite{Oe98}Oka et al. (1998).
They identified $\gsim 15$ molecular clouds with
a size of $\sim 30$~pc and obtained a total molecular mass in all clouds of
$\sim 4 \times 10^7 \, M_\odot$.  We take 20 for the number of GMCs,
$N_{GMC}$, and $2.5 \times 10^7$ or $5 \times 10^7 \, M_\odot$
for the total mass in GMCs, $M_{GMC}$.  For the sake of simplicity, a Plummer
model is adopted for the potential by the GMC:
\begin{equation}
\label{plummer}
	\Phi_{GMC} = -{G m_{GMC} \over (d^2 + \epsilon_{GMC}^2)^{1/2}},
\end{equation}
where $m_{GMC}$ is the mass of a GMC, $\epsilon_{GMC}$ the core radius
of the Plummer model, and $d$ the distance to the GMC.  Considering the
size of the GMCs in the inner bulge, we use $\epsilon_{GMC}$ of 5 or 10~pc.

Since only the projected positional distribution and the line-of-sight velocity
of the GMCs are available, the three-dimensional positional and velocity
distribution of GMCs must be assumed.  We assume that the $R$ (galactocentric
radius projected onto the Galactic plane; note that we use $r$ for the
three-dimensional galactocentric radius and $l$ for the projection of $R$
onto the plane of the sky) distribution ranges from 10 to 150~pc
and that the surface number density varies as $R^{\alpha_{GMC}}$.
\markcite{Be91}Binney et al. (1991) proposed that GMCs inside the
180-pc molecular ring move along the so-called X$_2$ orbits
(\markcite{CM77}Contopoulos \& Mertzanides 1977) and are gradually
transported inward at a constant velocity due to angular momentum loss.
Since the gas is thought to flow inwards along the Galactic
plane, if the gas is transported inward while conserving its
mass, $\alpha_{GMC}$ is required to be $-1$.  We here try $\alpha_{GMC}=-1$
and 0.  The inward migration takes place on a time scale much longer than
that of the X$_2$ orbital motion, thus we neglect the inward migration and
assume that the GMC motions in the plane follow X$_2$ orbits.  We also
assume that the CMZ maintains the same properties over the entire computational
time interval.

The information on the motion of GMCs along the $z$-axis and its
dependence on $R$ are very limited.
Therefore we assume that the vertical velocity dispersion of the GMCs,
$\sigma_{v,z}^{GMC}$, is constant over $R$ and simply impose a sinusoidal
vertical motion to the GMCs with an amplitude $H_{GMC}(R)$ that corresponds
to the adopted $\sigma_{v,z}^{GMC}$.  Under the standard background potential
adopted below, the X$_2$ orbital motion with $\sigma_{v,z}^{GMC}=25 \,
{\rm km\,s^{-1}}$ gives $H_{GMC}$ nearly proportional to $R$,
with $H_{GMC} \simeq 25$~pc at $R=150$~pc.
We use $H_{GMC}=25 \, (R/150{\rm pc})$~pc in the simulations.  
For comparison, the 15 giant molecular
clouds observed by \markcite{Oe98}Oka et al. (1998) have an average
$z$-distance of 18~pc, and \markcite{Oe98}Oka et al. estimate
$\sigma_{v,z}^{GMC}$ to be $\gsim 32 \, {\rm km \, s^{-1}}$.
The initial positions of the GMCs are randomly chosen, but
follow the assumed radial distribution, $R^{\alpha_{GMC}}$, and GMCs are
assumed not to interact with each other.  In order to lessen the CPU burden,
we pre-calculate the X$_2$ orbit of each GMC and save it in a table.
Then, when the equation of motion for the stars is integrated, the $x$-$y$
position of each GMC at a given time is obtained from the table and
the $z$ position is obtained from the sinusoidal motion with period
$4 H_{GMC}/\sigma_{v,z}^{GMC}$, and with $\sigma_{v,z}^{GMC}
= 25 \, {\rm km \, s^{-1}}$.

\subsection{Background Potential}
\label{sec:potential}

We model the potential of the inner bulge as
\begin{mathletters}
\begin{eqnarray}
	\label{phi_a}
	\Phi & =       & \Phi_0 \left ( 1 + \frac{a}{a_0} \right )^{2+\alpha}
			 - \frac{M_{bh}}{(r+r_0)} \\
	\label{phi_b}
	a    & \equiv  & \left  ( x^2+\frac{y^2}{y_0^2}+\frac{z^2}{z_0^2}
			 \right )^{1/2}
\end{eqnarray}
\end{mathletters}
where $r$ is the galactocentric radius and the softening parameter $r_0$,
which has negligible effect on the results, is adopted simply to avoid
numerical difficulties.  The first term in the right-hand-side of
equation~(\ref{phi_a}) is a softened power-law ellipsoidal potential,
and the second term, negligible at radii beyond a few parsecs,
is to account for the contribution to the potential
by the central black hole.  One could obtain a more complicated
potential model based on the the density model that matches the observed
luminosity profile (e.g., one based on an ellipsoidal density distribution),
but the simple potential above has been adopted here for the sake of efficient
force calculation.

The ellipsoidal potential is assumed
to rotate with a pattern speed $\Omega_p = 50$ or $65 \, {\rm km \, s^{-1} \,
kpc^{-1}}$, following \markcite{M90}Menzies (1990),
\markcite{Be91}\markcite{BGS97}Binney et al. (1991, 1997),
\markcite{ZRB96}Zhao et al. (1996), and \markcite{D99}Dehnen (1999).
Of course, the potential in the outer bulge will not be well described
with the above formula, but few inner bulge stars will have enough
kinetic energy to reach the outer bulge.  The masses in the outer bulge
and the Galactic disk presumably have a non-negligible contribution to
the potential in the inner bulge, but such contributions
may be understood to be already included in parameters $y_0$ and $z_0$
in equation~(\ref{phi_b}).  While \markcite{BN68}Becklin \& Neugebauer (1968)
found the IR luminosity distribution for $r \leq 25$~pc to be $\sim r^{-1.8}$,
the kinematic study by \markcite{LHW92}Lindqvist, Habing, \& Winnberg (1992a)
implies a density distribution in the $30 \, {\rm pc} \lsim r \lsim
100 \, {\rm pc}$ region proportional to $\sim r^{-1.5}$ (for $1 \, {\rm pc}
\lsim r \lsim 100 \, {\rm pc}$, the profile becomes $\sim r^{-2.0}$).
We use $\alpha = -1.8$ or $-1.5$.  For both $a_0$ and $r_0$, we
adopt 1~pc.  The fit to the observations with the dust model of
\markcite{SMB96}Spergel et al. (1996) gives a density aspect ratio of 1.67
(\markcite{BGS97}Binney, Gerhard, \& Spergel 1997).  We find that our
potential model with $z_0=0.8$ gives a density aspect ratio of $\sim 1.7$.
We thus choose $z_0=0.8$ and 0.7.
For the set of parameters adopted here, the transition from
X$_1$ orbits to X$_2$ orbits, which \markcite{Be91}Binney et al. (1991)
propose to be responsible for the 180-pc molecular ring, forms at
$r \simeq 180$~pc only when $y_0 \geq 0.9$.  Thus we adopt $y_0=0.9$.
Finally, we choose $\Phi_0$ such that $\Phi_0 (1+r/a_0)^{2+\alpha}$
corresponds to a mass distribution with a total mass inside $r=30$~pc
of $8 \times 10^7 \, M_\odot$ (all models except model~4;
\markcite{LHW92}Lindqvist et al. 1992a) or $2 \times 10^8 \, M_\odot$
(model~4; \markcite{GT87}Genzel \& Townes 1987).
The equipotential and isodensity contours of the above model are
shown in Figure~\ref{fig:potential}.  The density distribution of a model
with $z_0=0.7$ is slightly pinched at the z-axis (perpendicular to the
Galactic plane).  However, the calculation actually uses the derivative of
the potential, and neither the potential nor its derivative has such a feature.
Thus, the pinch in the density will not significantly affect our results.

\subsection{Test Particles}
\label{sec:particles}

At every 10~pc between radii of 10 and 150~pc,
300 test particles (stars) are initially distributed in the plane around
a ring of that radius, making a total of 4500 stars per simulation.
This discrete initial distribution allows us to apply one simulation
result to initial power-law distributions of stars with various exponents
by differently weighting the stars initially placed at different $R$.
Simulation results are analyzed here for $\alpha_* = -2$ and $-1$,
where $\alpha_*$ is the exponent of the initial, power-law surface
density profile of stars in the plane ($R^{\alpha_*}$ is a surface
density profile projected along the $z$-axis, not along the line-of-sight).
The stars' initial azimuthal angles are random.  Stars initially move on
X$_2$ orbits, and Gaussian random motions parallel and perpendicular
to the plane are imposed, with standard deviations $\sigma_{v,xy}=10$--$20
\, {\rm km \, s^{-1}}$ and $\sigma_{v,z}=12$--$25 \, {\rm km \, s^{-1}}$,
respectively.  The value of $25 \,{\rm km \, s^{-1}}$ for $\sigma_{v,z}$
is chosen to match the $\sigma_{v,z}^{GMC}$ value introduced earlier in
deriving $H_{GMC}=25\, (R/150{\rm pc})$~pc.  Stars are first located at
$z=0$ and then are propagated for approximately 10 orbital
periods to make the initial distribution phase-mixed.  Since
the dynamical relaxation time in the inner bulge is larger than
the Hubble time, we neglect the interaction between stars.

\subsection{Numerical Method}
\label{sec:method}

We have tested three numerical methods for integrating the equation
of motion: the adaptive step size Runge-Kutta method, the Richardson
extrapolation method (Bulirsch-Stoer method, \markcite{Pe92}Press
et al. 1992), and the predictor-corrector method (variable time step
leapfrog method; \markcite{HMM95}Hut, Makino, \& McMillan 1995),
and found that the Bulirsch-Stoer method in a rotating frame is
the most efficient for our problem.  We also tried Stoermer's
scheme, which is for second-order conservative equations like
the equation of motion without a derivative on the right-hand-side
(see, e.g., \markcite{Pe92}Press et al. 1992), in conjunction with
the Bulirsch-Stoer method, but found it to be slightly less efficient
than our choice above.
In our simulations, the energy of a single star is conserved to an
accuracy of better than $3 \times 10^{-6}$ during the entire time
interval followed.

\begin{figure*}[p]
\parbox{8.7cm}{
%Fig 2
\centerline{\epsfxsize=16cm\epsfbox{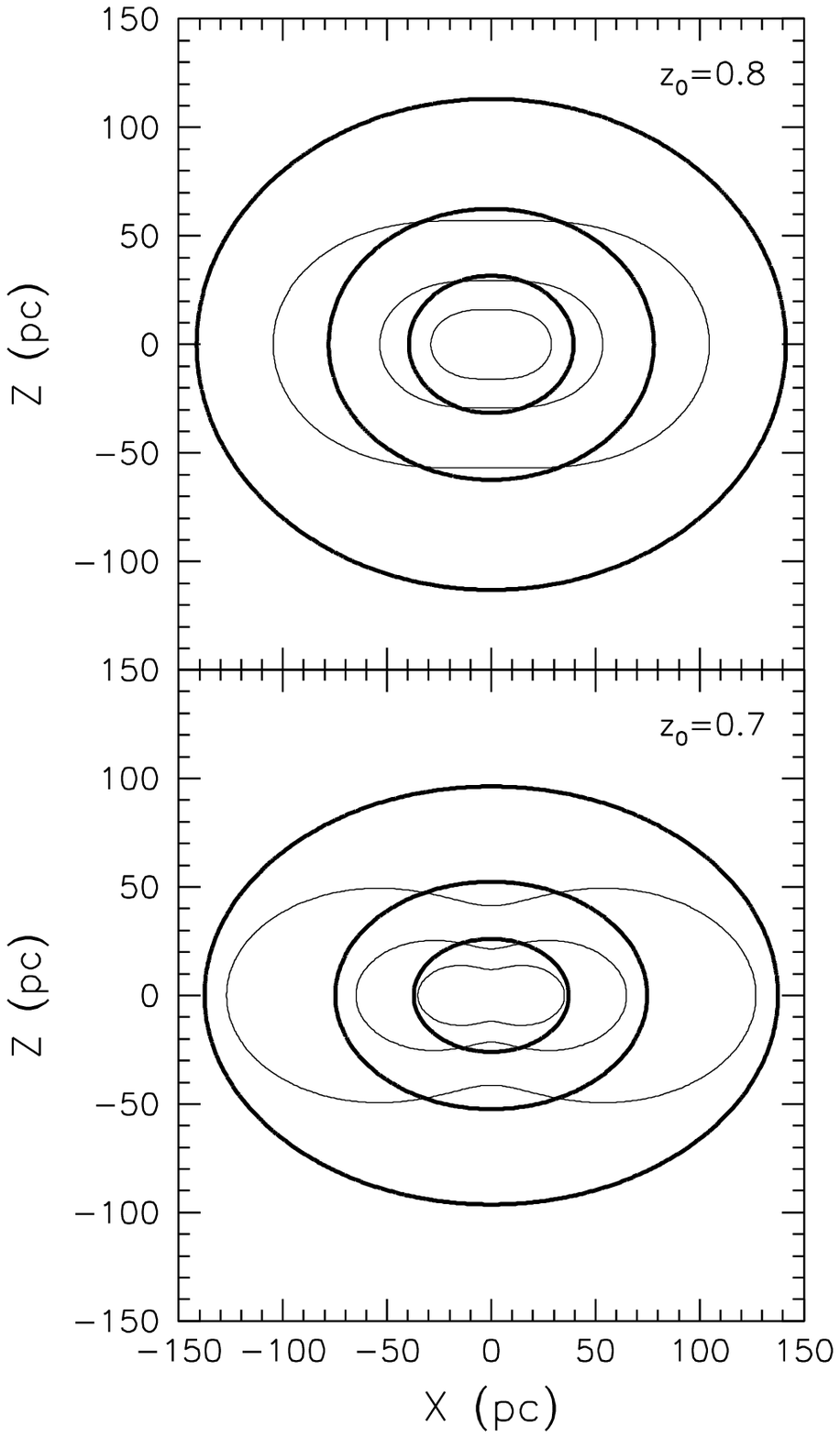}}
\caption
{\label{fig:potential}
Contours of equipotential ({\it thick lines}) and isodensity
({\it thin lines}) for our bulge potential model with $z_0=0.8$
(model~1 [our standard model]; {\it upper panel}) and $z_0=0.7$
(model~2; {\it lower panel}).  Contours are drawn at linear steps
for the potential and logarithmic steps for the density.  The x-axis
lies in the Galactic plane and the z-axis is perpendicular to the plane.
}
%\end{figure*}
}
\hfill
\parbox{8.7cm}{
%\begin{figure*}
%Fig 3
\centerline{\epsfxsize=8.8cm\epsfbox{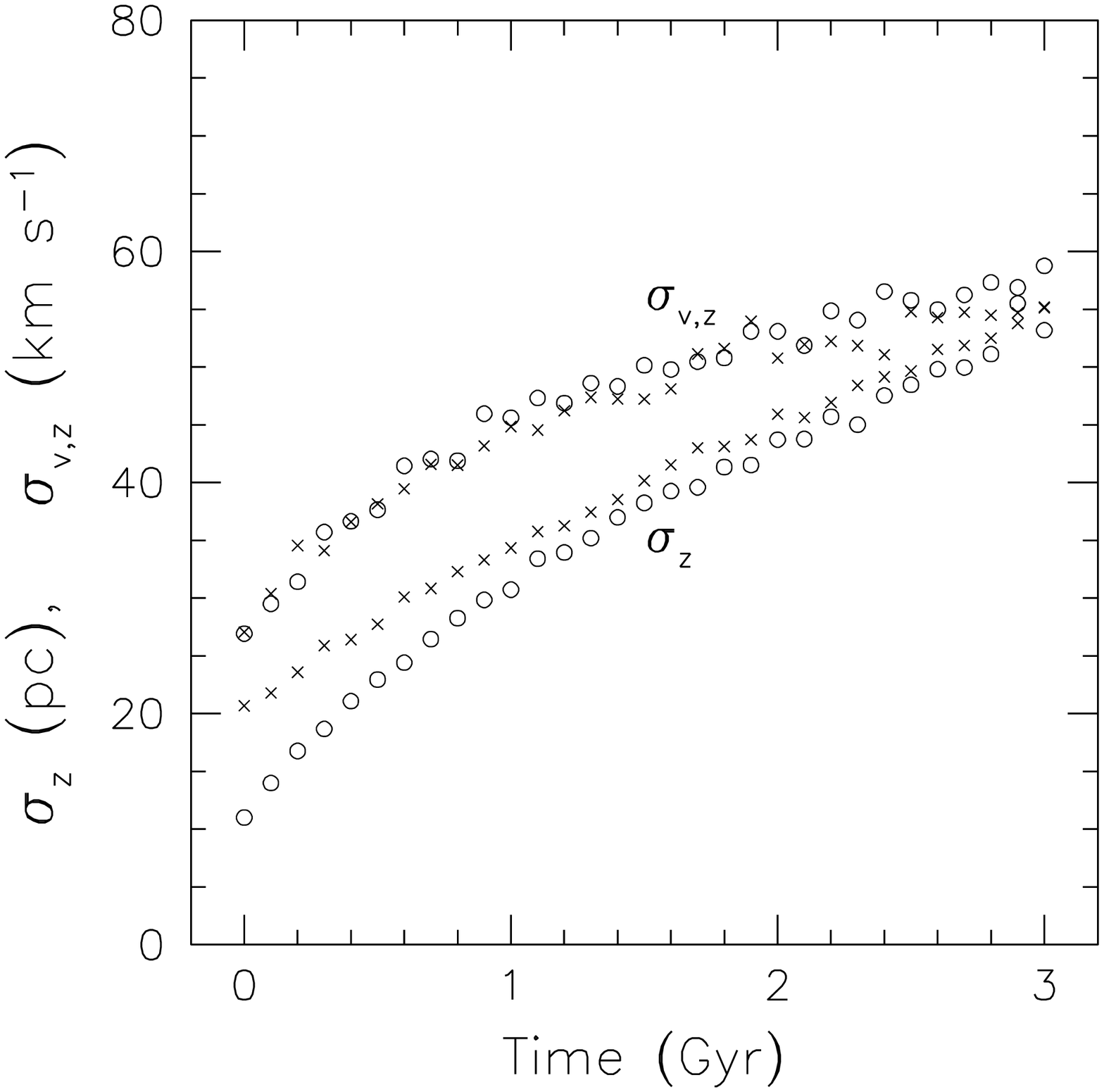}}
\caption
{\label{fig:sigg1} Evolution of dispersions of z-axis stellar
positions and velocities for model~1.  Stars initially located at
$30 \le R {\rm /pc} \le 70$ ({\it circles}) and $80 \le R {\rm /pc}
\le 120$ ({\it $\times$'s}) are separately presented.
}
%\end{figure*}

\vspace{1.5cm}
%\begin{figure*}
%Fig 4
\centerline{\epsfxsize=8.8cm\epsfbox{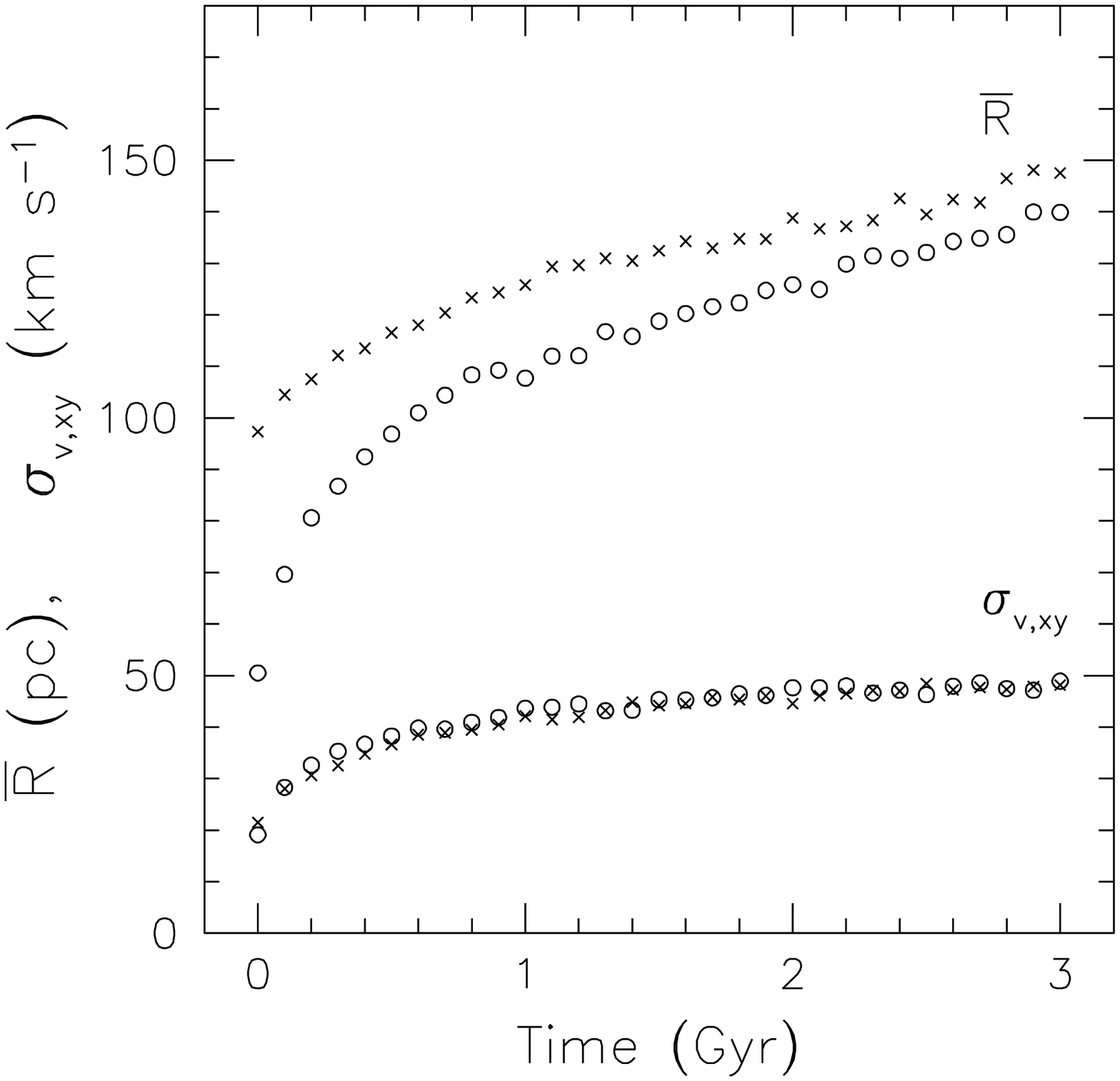}}
\caption
{\label{fig:xy} Evolution of the average $R$ values and the velocity
dispersion in the plane for model~1.  Symbols have the same meaning as
in Figure~\ref{fig:sigg1}.
}
}
\end{figure*}

%%%%%%%%%%%%%%%%%%%%%%%%%%%%%%%%%%%%%%%%%%%%%%%%%%%%%%%%%%%%%%%%%%%%%%%%%%%%%%%%
\section{RESULTS}
\label{sec:results}

\subsection{Scale Heights and Velocity Dispersions}
\label{sec:velocity}

Figure~\ref{fig:sigg1} shows the 3~Gyr evolution of the
dispersions of vertical stellar positions and velocities for
model~1.  Stars initially at $30 \leq {\rm R/pc} \leq 70$ and
at $80 \leq {\rm R/pc} \leq 120$ are separately shown.
Both vertical heights and velocities significantly increase
in the first 1~Gyr.  The initial $\sigma_z$ values are smaller
for the group of stars with smaller $R$ because the adopted potential
for the bulge has a larger absolute vertical gradient at smaller
$R$.  It is interesting that the increase of vertical velocity is almost
independent of $R$, since stars closer
to the Galactic center are expected to encounter GMCs more frequently
than the ones at larger distances for this model (rotational velocity
is nearly independent of $R$ and the surface density of GMCs is larger
at smaller $R$ for this model) and thus one would expect the velocities
of stars closer to the center to grow faster.
We attribute this independence of vertical velocity evolution on $R$ to the
rapid evolution of the motions in the plane for stars at smaller
$R$.  Figure~\ref{fig:xy} shows the average radii in the plane, $\bar R$, for
two radial groups of stars of model~1.  $\bar R$ of the stellar group with
initially smaller $R$ doubles and becomes comparable to that of the group
with initially larger $R$ in only 0.5~Gyr.  Furthermore, we find that
the average ratio of apocenter to pericenter radii of stellar orbits rapidly
increases from $\sim 1.5$ to $\sim 4$ in 0.5~Gyr for both groups, and keeps
increasing afterwards.  Once the orbit of a star becomes highly
eccentric, the star may encounter any GMC
in the CMZ during its orbital motion, regardless of its initial location
in the CMZ (or, regardless of its initial local number density of GMCs).

Also shown in Figure~\ref{fig:xy} is the evolution of the velocity
dispersion in the plane, $\sigma_{v,xy}$, for model~1.
Unlike $\sigma_{v,z}$, $\sigma_{v,xy}$ ends its rapid increase and
flattens after $t \simeq 1$~Gyr.  The radial velocity
dispersion in the plane, $\sigma_{v,R}$, and the tangential velocity
in the plane, $\sigma_{v,\theta}$, have comparable values
($0.9 \lsim \sigma_{v,\theta}/\sigma_{v,R} \lsim 1.1$),
implying that the velocity dispersion in the plane is nearly isotropic.
We also find that $\sigma_{v,z}/\sigma_{v,R}$ stays mostly between 0.7
and 0.8, which is slightly larger than the value of 0.6 found by
\markcite{V85}Villumsen (1985) for the Galactic disk case.

\begin{figure*}
%Fig 5
\centerline{\epsfxsize=17cm\epsfbox{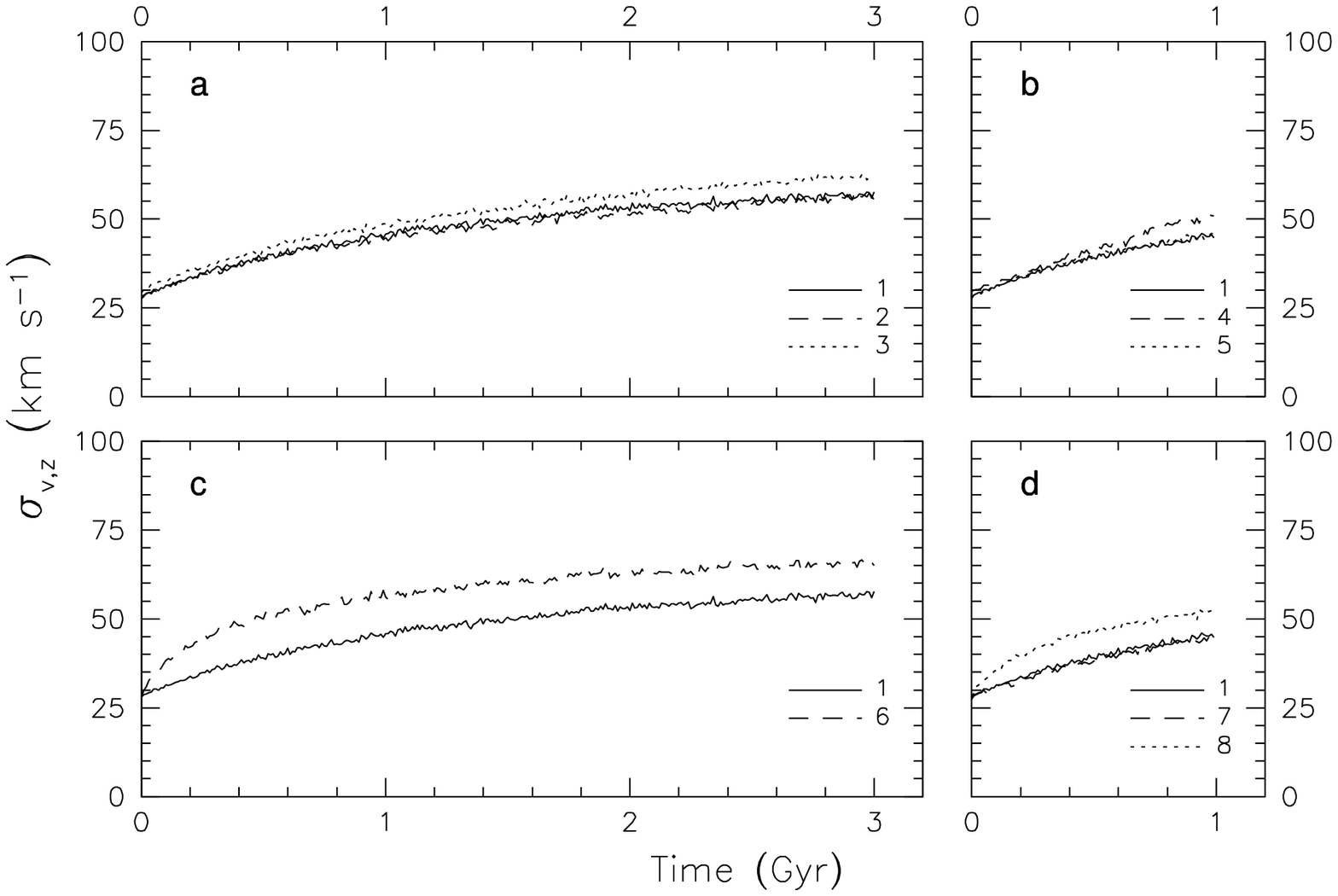}}
\caption
{\label{fig:allsig} Evolution of vertical velocity dispersion.
a) models 1, 2, and 3 (different Galactic potential models);
b) models 1, 4, and 5 (different bulge masses and pattern speeds);
c) models 1 and 6 (different GMC masses);
and d) models 1, 7, and 8 (different GMC distribution and sizes).
}
\end{figure*}

The growth of the velocity dispersion may be described by the diffusion
equation
\begin{equation}
\label{sigvm}
	{d \sigma^2 \over dt} \propto \sigma^{-m},
\end{equation}
whose solution has a functional form of
\begin{equation}
\label{sigvn1}
	\sigma(t) = \sigma_0 \left (1+{t \over \tau} \right )^n,
\end{equation}
with $n=1/(m+2)$.  When $t \gg \tau$, equation~(\ref{sigvn1}) simplifies to
\begin{equation}
\label{sigvn2}
	\sigma(t) \propto t^n.
\end{equation}
\markcite{V85}Villumsen (1985) found that the vertical velocity dispersion
of stars in the Galactic disk is well fit by equation~(\ref{sigvn1}) with
$n=0.38$ ($m=0.63$) for the first 1~Gyr, and by equation~(\ref{sigvn2})
with $n=0.31$ ($m=1.2$) for $t>0.3$~Gyr.  These values of $n$ are only
slightly larger than the theoretical estimates by
\markcite{L84}Lacey (1984), 0.33 and 0.25, respectively.  The theoretical
estimation of these values for the inner Galactic bulge, however, is not
easy because 1) the background potential is not plane-parallel,
2) the orbits of stars may not be approximated as epicyclic motions,
and 3) the influence of GMCs may not be independent of each other because
of their large number density and softening radius, $\epsilon_{GMC}$.
Nonetheless, a comparison of the growth rate for the bulge and the disk may
be instructive.  We applied non-linear, 3-parameter least squares fits
(eq.~[\ref{sigvn1}]) to the early phase of our vertical velocity dispersion,
and linear, 2-parameter least squares fits (eq.~[\ref{sigvn2}]) to the later
data, and find that the growth rates 
depend on the choice of time range to fit.  Model~1 shows $n=0.30$
($m=1.33$) for $t \leq 1$~Gyr and $n=0.20$ ($m=3.0$) for $t \geq 1$~Gyr,
and model~3 gives $n=0.32$ ($m=1.1$) and $n=0.23$ ($m=2.3$), respectively.
As in the Galactic disk, $n$ becomes smaller as the scattering off the GMCs
drives stars farther and farther away from where the GMCs reside.

Next, we compare the evolution among different models.
The velocity evolution of all our models is shown in
Figure~\ref{fig:allsig}.  Panel a) shows simulations with different
bulge potential models, b) with different bulge masses inside 30~pc and
pattern speeds, c) with different GMC masses, and d) with different GMC
distributions and sizes.  Panels a) and b) show that in general, the vertical
velocity evolution is not sensitive to the parameters used to model the bulge.
The heating is only slightly more efficient when the potential is
shallower (larger $\alpha$).
%It is conceivable that the
%relatively smaller scale height, which keeps stars nearer to GMCs, is
%responsible for the faster heating of stars, at a given $\sigma_{v,z}$.

On the other hand, panels c) and d) of Figure~\ref{fig:allsig} illustrate
that the vertical velocity evolution is dependent on the mass and size of
GMCs, but not on the distribution of GMCs.  Stars are found to have highly
eccentric orbits, which allows them to encounter any GMC in the
CMZ.  Thus the heating rate is determined mostly by the total number and mass
of the GMCs in the CMZ, and not by the distribution of GMCs.

Although not shown in Table~\ref{table:runs} and
Figure~\ref{fig:allsig}, we have tried other initial
stellar velocity dispersions ($\sigma_{v,xy}=10 \, {\rm km \, s^{-1}}$ and
$\sigma_{v,z}= 12 \, {\rm km \, s^{-1}}$), but found no significant
difference from our standard model.  The velocity dispersions rapidly
increase by a factor of a few in the early phase, so unless the initial
velocities are different by an order of magnitude, any initial difference
in velocity dispersions that is smaller than the amount of increase in the
early phase will soon be indistinguishable.

\subsection{Density Profiles and Aspect Ratios}
\label{sec:aspect}

In this subsection, we discuss quantities with which we may more directly
assess the hypotheses that the $r^{-2}$ cluster is a result of sustained
star formation in the CMZ and that the GMCs are responsible for the vertical
diffusion of newborn stars.  To do so, we project the simulation stars onto
the $l$--$b$ plane of the sky and compare them with observations of a certain
stellar type.  From a variety of models of the inner Galaxy, the major axis
of the bulge bar potential has been found to
have an angle of $15$--$30\arcdeg$ to the east from the line connecting
the Galactic center and the Sun (\markcite{BGS97}Binney et al. 1997,
among others).  Here we adopt $20\arcdeg$.

\begin{figure*}
%Fig 6
\centerline{\epsfxsize=16cm\epsfbox{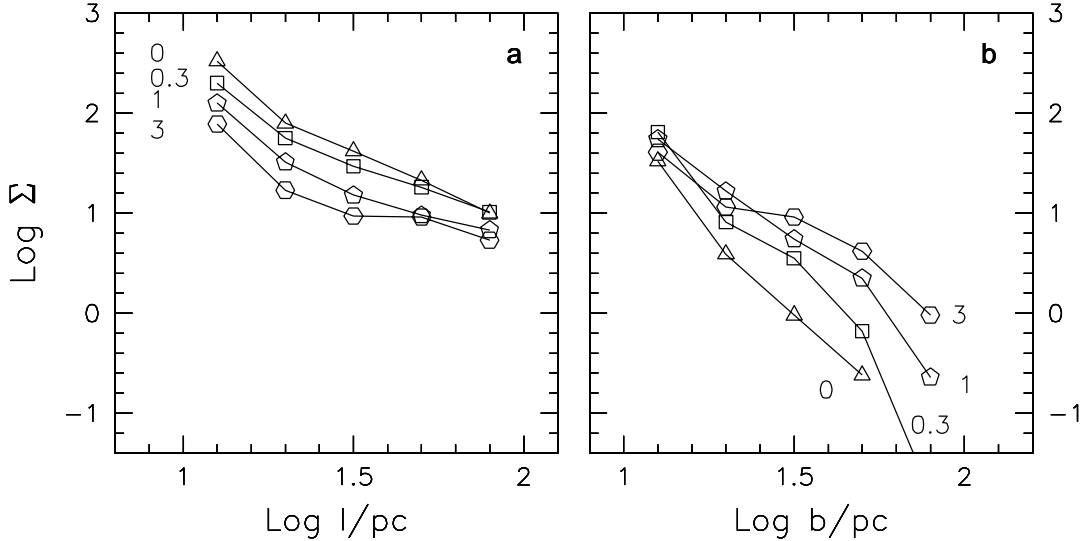}}
\caption
{\label{fig:prof} Surface density profiles (projected along the line-of-sight)
for model~1
along the $l$ axis ({\it a}) and along the $b$ axis ({\it b}) at four
epochs, 0, 0.3, 1, and 3~Gyr.  Stars that are less than 10~pc away from
the axes are considered.  The numbers in the plot denote
the time in Gyr.  $\Sigma$ is in arbitrary units.
The angle between the major axis of the bar and
the Sun-Galactic center line is assumed to be $20 \arcdeg$, and
$\alpha_* = -2$ is adopted.  The 1-$\sigma$ Poisson error in log~$\Sigma$
is smaller than 0.05 octave except for Log~$\Sigma < 0$.
}
\end{figure*}

Our simulations initially have an equal number of stars for each linear
$R$ bin, making the exponent of the radial profile of the initial
surface density of simulation stars in the plane, $\alpha_*$, equal to $-1$
(note that $\alpha_*$ is different from $\alpha$, the exponent for the bulge
density profile).  However, by properly weighting
the stars in each bin, one may construct volume and projected
(along the line-of-sight) surface
density profiles (we denote the latter with $\Sigma$) for different
$\alpha_*$ values.  Figure~\ref{fig:prof} shows the $\Sigma$ profile
along the $l$ and $b$ axes at several epochs for $\alpha_*=-2$.
It is interesting that $\Sigma$ along the $l$ axis gradually decreases
at $l<30$~pc while maintaining its overall profile.
The $\Sigma$ profile along the $b$ axis experiences a significant flattening
in its power-law slope.

\begin{figure*}
%Fig 7
\centerline{\epsfxsize=8.8cm\epsfbox{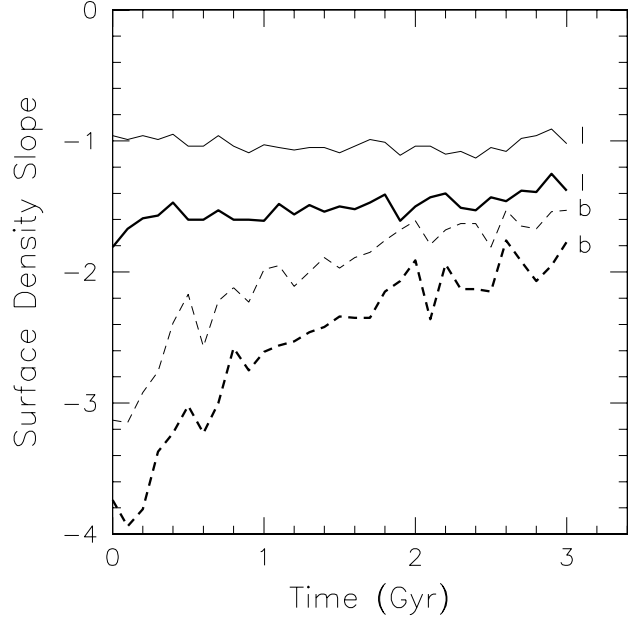}}
\caption
{\label{fig:axis} Evolution of the surface density slope (projected along
the line-of-sight)
for model~1 along the $l$ axis ({\it solid lines}) and along the
$b$ axis ({\it dashed lines}).  The slopes are the best-fit power-law
exponent to the distributions obtained in Figure~\ref{fig:prof}.
The slopes are calculated for the $\alpha_* = -2$ case ({\it thick lines})
and the $\alpha_* = -1$ case ({\it thin lines}).
}
\end{figure*}

Plotted in Figure~\ref{fig:axis} is the evolution of the best-fit
power-law exponent for the $\Sigma$ profile between 10 and 100~pc
along the $l$ and $b$ axes.  As anticipated from Figure~\ref{fig:prof},
while the $l$ axis slope does not vary much, that along the $b$ axis
continuously increases and approaches the former at the
end of the simulation.  The slope values for the $\alpha_*=-2$ case are steeper
than those for the $\alpha_*=-1$ case by $\sim 0.5$ (when the distribution
extends to infinity, a difference of 1 in $\alpha_*$ should result
in a difference of 1 in the power-law $\Sigma$ slopes;  however, the
stellar distribution in our simulations, as in reality, is finite,
so the relation between $\alpha_*$ and the $\Sigma$ slope is not as
strong as in the infinite case).  Using a circular
aperture, \markcite{BN68}Becklin \& Neugebauer (1968) obtained
integrated near-infrared intensities of the inner Galactic bulge region
as a function of aperture size and derived a surface intensity profile
(projected along the line-of-sight)
proportional to $r^{-0.8}$.  The enclosed mass in the $5 \lsim r/{\rm pc}
\lsim 100$ region obtained by \markcite{LHW92}Lindqvist et al. (1992a) also
implies a surface density profile of $\propto r^{-0.8}$.  On the other hand,
by fitting an ellipse with an aspect ratio of 2.2 to the distribution
of stars with dereddened, apparent K mag between 5 and 6
(\markcite{HR89}Haller \& Rieke 1999 estimate such stars
to be a few $10^8$~yr old),
\markcite{CWG90}Catchpole et al. (1990) obtained a steeper exponent,
$-1.4$, after correcting for the disk contamination, crowding, and dark clouds.
The invariance of the $l$-axis slope over time
in our simulations makes the $l$-axis slope indicative of its original slope.
From Figure~\ref{fig:axis}, we find that the deprojection of our simulation
with $\alpha_*=-1$ approximately reproduces the slope obtained by Becklin
\& Neugebauer and Lindqvist et al., and that with $\alpha_*=-2$ gives the
slope obtained by Catchpole et al.
%\markcite{SM96}Serabyn \& Morris (1996) speculate two possible origins
%of a star formation efficiency resulting in $\alpha_*=-2$.  One involves
%the collision of molecular clouds having a distribution $\propto R^{-1}$,
%which is presumably the result of conservation of the mass inflow in the
%disk.  The other involves a rotating spiral pattern that sweeps through
%material distributed as $\propto R^{-1}$ at a frequency $\propto R^{-1}$.
In any case, it appears reasonable to assume that the time-averaged
star formation efficiency in the inner bulge has a dependence between
$R^{-2}$ and $R^{-1}$ in the plane.

Finally, we compare the vertical thickening of our simulation stars with
the observations.  Since it is more instructive to compare with
the observation of a coeval population (otherwise, the overlap of
populations of different ages would blend out the age dependence
of a certain quantity), we use the OH/IR stars observed by
\markcite{Le92}Lindqvist et al. (1992b) in the inner bulge
(total number of 134).  One of the
measures of the thickness or flatness of a distribution is the
aspect ratio of major ($l$) to minor ($b$) axes.  However, the aspect
ratio itself can be defined in several ways.  \markcite{CWG90}Catchpole
et al. (1990) measured the aspect ratio of the stellar distribution using
observed isodensity contours.  However, we find that a distribution of
134 stars is too sparse to obtain meaningful contours.
For this reason, we determine the aspect ratio as the ratio of the major
to minor axes of an ellipse which best fits the distribution of stars
projected onto the plane of the sky.  The best-fit ellipse was taken to
be that for which the area per unit angle, taken as a function of angle
on the sky measured from the center of the ellipse, was best matched by
the function describing the number of stars per unit angle.  The binning
utilized to determine the latter function was chosen to ensure a
statistically large number of stars in each angular bin.  The orientation
of the major axis of the fitted ellipse was fixed at the orientation of
the Galactic plane.  This is a one-dimensional fit, thus is less sensitive
to the density profile, which is not a primary concern here.
The observation by Lindqvist et al. was constrained
to a cross-shaped area, but we limit the calculation of the aspect
ratio of both observations and simulations to the stars in the
$200 \, {\rm pc} \times 80 \, {\rm pc}$ area positioned at the Galactic center
(the number of OH/IR stars observed by Lindqvist et al. in this area is 124).
We here assume $\alpha_*=-2$, but we find that $\alpha_*=-1$ gives nearly
the same aspect ratios.

\begin{figure*}
%Fig 8
\centerline{\epsfxsize=17cm\epsfbox{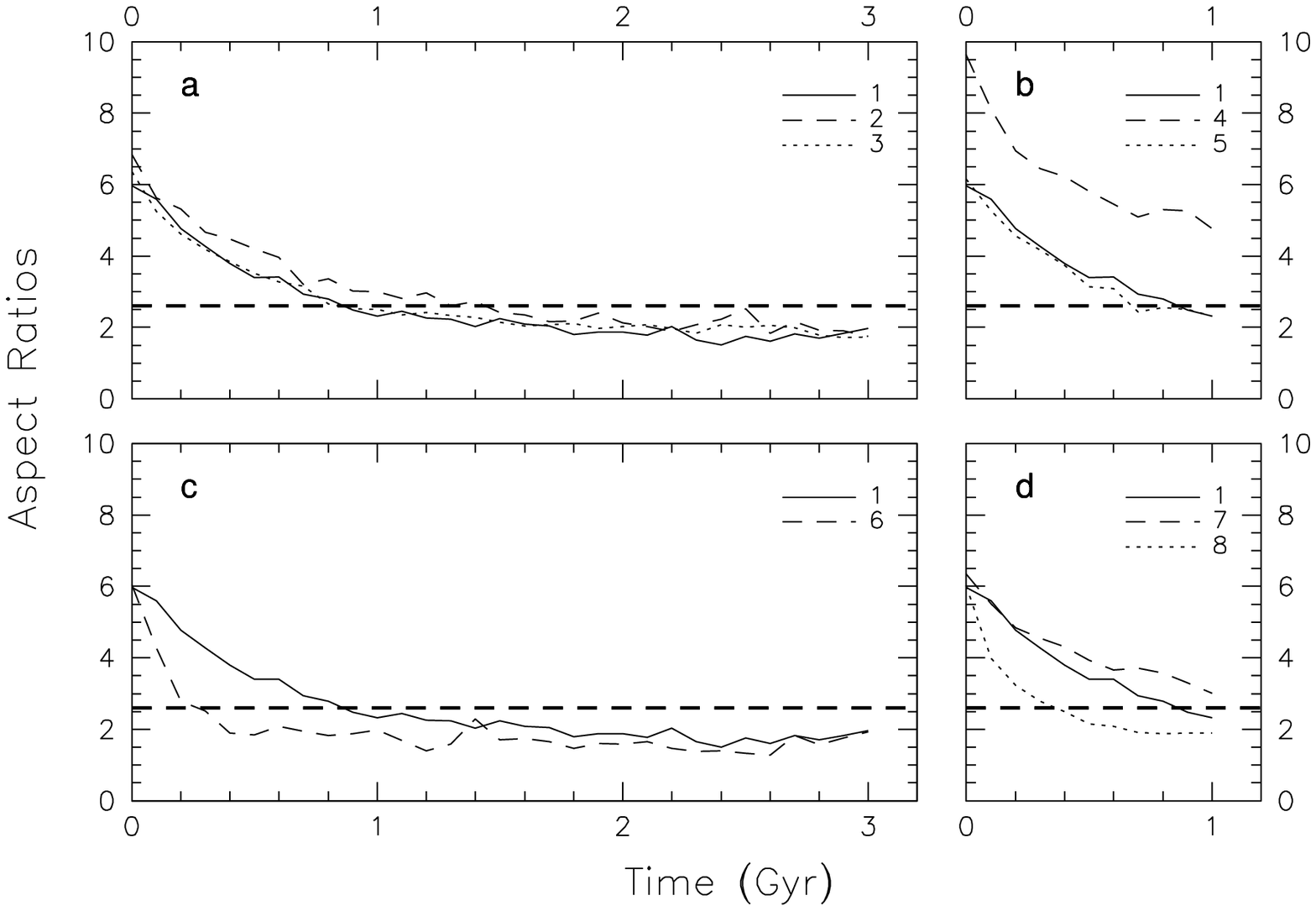}}
\caption
{\label{fig:allaspect} Evolution of aspect ratios of our models.
See the caption of Figure~\ref{fig:allsig} and Table~\ref{table:runs}
for the description of each panel.  See text for our definition of
aspect ratio.  The horizontal dashed line indicates the aspect ratio
from the distribution of OH/IR stars observed by Lindqvist et al. (1992).
$\alpha_* = -2$ is adopted.  The 95~\% confidence limit for the
observed aspect ratio is 2.6$^{+0.5}_{-0.5}$.
}
%\end{figure*}

\vspace{0.5cm}
%\begin{figure*}
%Fig 9
\centerline{\epsfxsize=17cm\epsfbox{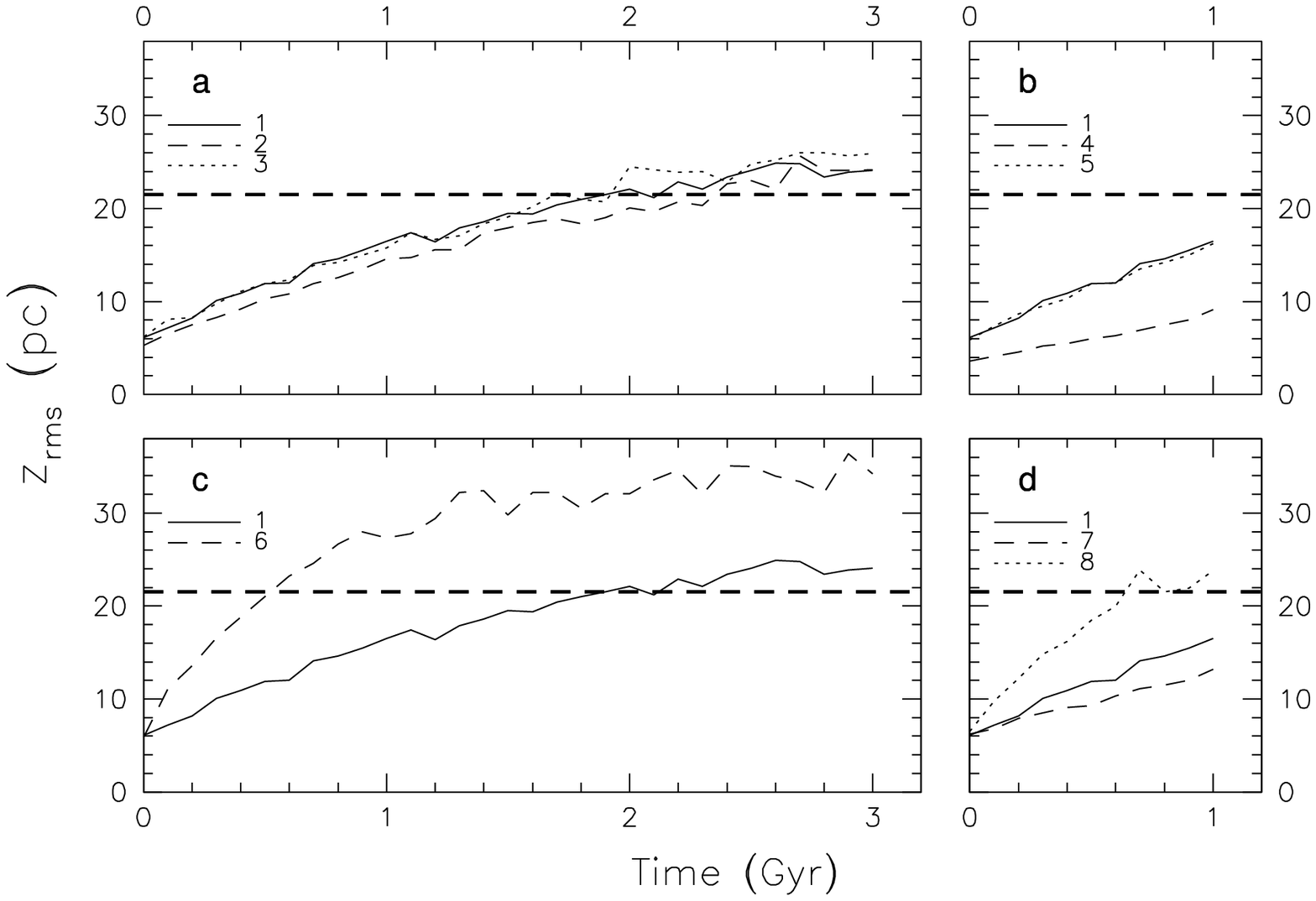}}
\caption
{\label{fig:allzrms} Evolution of the root-mean-square value of the Galactic
height, $z_{rms}$, from our simulations.  The calculation of $z_{rms}$ is
limited to stars with $|l|<20$~pc and $|b|<80$~pc.
See the caption of Figure~\ref{fig:allsig}
and Table~\ref{table:runs} for the description of each panel.  The horizontal
dashed line indicates the $z_{rms}$ value from the distribution
of OH/IR stars observed by Lindqvist et al. (1992).
$\alpha_* = -2$ is adopted.  The 2--$\sigma$ uncertainty level for
$z_{rms}$ values is less than 10~\%.
}
\end{figure*}

Figure~\ref{fig:allaspect} shows the evolution of the aspect ratios of
all models.  The evolution is noticeably dependent only on the mass
and size of GMCs as well as the bulge mass.  The former dependence is
a naturally expected phenomenon, and the latter dependence appears to be
due to an initially larger aspect ratio of model 4, which has a larger
vertical potential gradient, but the same initial vertical velocity
dispersion as the other models.

In spite of the dependence on the GMC parameters, all models except
model~4 (larger bulge mass case) evolve to having aspect ratio values
below 3 in 1--$2$~Gyr.  The same elliptical fit to the distribution of
observed OH/IR stars in the central $200 \, {\rm pc} \times 80 \, {\rm pc}$
area gives an aspect ratio of 2.6$^{+0.5}_{-0.5}$ (super- and subscripts
denote 95~\% confidence limits), shown in Figure~\ref{fig:allaspect}
as a horizontal dashed line.  Based on the observed bolometric magnitudes of
the OH/IR stars ($M_{bol} \simeq -5$; \markcite{Je94}Jones et al. 1994
\& \markcite{Be98}Blommaert et al. 1998) and the theoretical initial
mass-bolometric magnitude relation of \markcite{VW93}Vassiliadis \& Wood
(1993), \markcite{Se99}Sjouwerman et al. (1999) estimates OH/IR stars
in the inner bulge to be 1--2~Gyr old.
Since the time scale for an initially
flattened stellar population to achieve an aspect ratio of the observed
OH/IR stars agrees well with the estimated age of the OH/IR stars,
we conclude that scattering by the GMCs is one of the predominant
mechanisms, perhaps the most important mechanism, for the vertical
diffusion of stars in the inner Galactic bulge.

It has been suggested by \markcite{Be81}Baud et al. (1981) that
the expansion velocity of the circumstellar shell of an OH/IR star,
$v_{exp}$, is a good statistical age indicator.  For the Galactic
inner bulge, \markcite{LHW92}Lindqvist et al. (1992a) confirmed
this correlation by showing that the OH/IR stars there
with $v_{exp} \geq 18 \, {\rm km \, s^{-1}}$
have younger kinematical and morphological characteristics
than the ones with $v_{exp} < 18 \, {\rm km \, s^{-1}}$.  Our aspect
ratio calculation for the same sample finds that such a division is
clearer when the division is made at $20 \, {\rm km \, s^{-1}}$:
while the aspect ratio of stars with $v_{exp} \leq 20 \, {\rm km \, s^{-1}}$
is 2.5$^{+0.5}_{-0.4}$, that of stars with $v_{exp} > 20 \,
{\rm km \, s^{-1}}$ is 3.2$^{+1.6}_{-1.3}$.  Although the latter,
which is calculated from a sample of only 29, is rather uncertain,
the rapid decrease of aspect ratios in our simulations supports
the notion that the former group is younger than the latter group.

Another measure for the comparison between our simulations and observations
is the root-mean-square (rms) value of the Galactic height ($z_{rms}$).
This is also a projected quantity, but is sensitive to the vertical
evolution of stars.  We limit the calculation of $z_{rms}$ to stars with
$|l|<20$~pc and $|b|<80$~pc, in order to minimize the effects of planar
evolution and of a few outliers with very large $|z|$ values.  The evolution of
$z_{rms}$ for all models is shown in Figure~\ref{fig:allzrms} along with
the $z_{rms}$ value for the OH/IR stars in the same region, 22~pc
(thick dashed lines).  The figure is made assuming $\alpha_*=-2$,
but we find that $\alpha_*=-1$ typically gives only 10--20~\% larger $z_{rms}$
values.  All models, probably except model 4, reach the
value for the OH/IR stars within $\sim 2$~Gyr, as in the case of the
aspect ratio evolution.  One noticeable behavior difference between
$z_{rms}$ and the aspect ratio is its largely different convergent values
between models 1 and 6.  Thus, $z_{rms}$ could be useful in comparing
simulations with the observed distribution of stars older than OH/IR stars.

\begin{figure*}
%Fig 10
\centerline{\epsfxsize=8.8cm\epsfbox{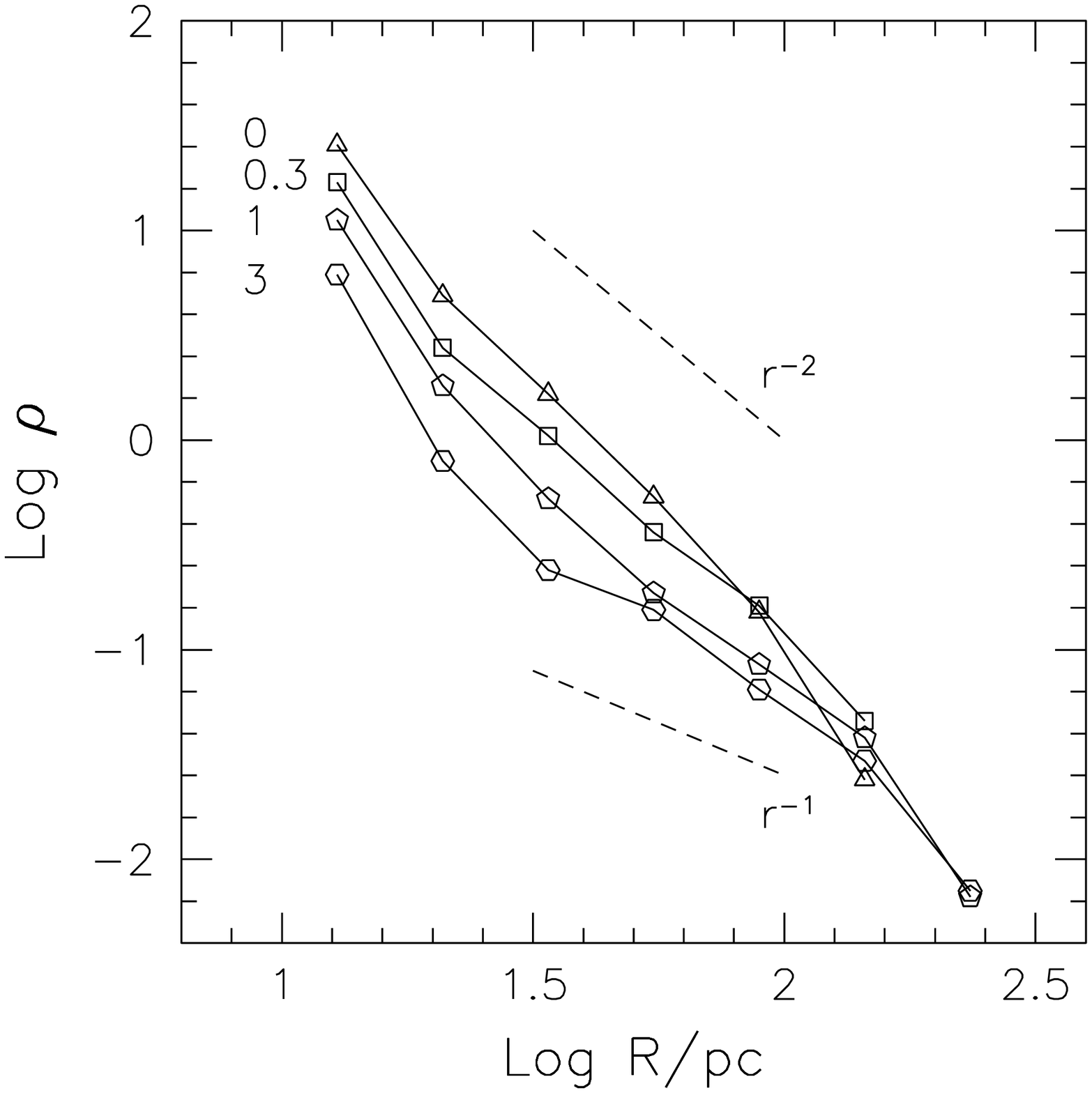}}
\caption
{\label{fig:rho} Volume density profiles in the plane, $\rho$, for model~1
at four epochs, 0, 0.3, 1, and 3~Gyr.  Stars that are less than 10~pc away
from the plane are considered.  The numbers in the plot denote the time
in Gyr, and dashed lines indicate profiles of $r^{-1}$ and
$r^{-2}$.  $\rho$ is in arbitrary units.  $\alpha_* = -2$ is adopted.  
The 1-$\sigma$ Poisson error is smaller than 0.05 octave.
}
\end{figure*}

%%%%%%%%%%%%%%%%%%%%%%%%%%%%%%%%%%%%%%%%%%%%%%%%%%%%%%%%%%%%%%%%%%%%%%%%%%%%%%%%
\section{DISCUSSION}
\label{sec:discussion}

The surface density profile along the $l$ axis of model~1 at 3~Gyr
(Fig.~\ref{fig:prof}a) shows a nearly flat slope near $l=30$~pc.
We find that this is caused by a relatively shallow volume density
profile in the plane between $r=30$~pc and 100~pc,
$\rho \sim r^{-1}$ (see Fig.~\ref{fig:rho}).  In the case of model~3
(shallower bulge potential), the volume density profile becomes even
shallower than $r^{-1}$ at 3~Gyr.  However, the profile of model~7,
where the GMC distribution is uniform over $R$, does not change much
over its simulation period, 1~Gyr.  Thus the shallow density profiles
shown in $\alpha_{GMC}=-1$ models are attributable to the relatively larger
abundance of GMCs at smaller $R$, which more efficiently depletes
stars in that region by heating.  It is difficult
to estimate the current volume density profiles in the inner bulge,
both observationally and from our simulations.  Observationally,
we only have projected surface density profiles and limited radial
velocity information.  To estimate the density profile from our simulations,
one needs to assume a star formation history and a radial dependence of
the star formation in the inner bulge.  Until these are better understood,
simulations may not give a strong constraint
to the density profile.  However, our simulations still seem to be able to
suggest that the local density in the plane between 30 and 100~pc may be
as shallow as or close to $r^{-1}$ (compared to $r^{-1.8}$ obtained by
\markcite{BN68}Becklin \& Neugebauer 1968 for $r<25\,{\rm pc}$
and $r^{-2.5}$ by \markcite{LHW92}Lindqvist et al. 1992a for
$1<r/{\rm pc}<10$).  Such shallow profiles may carry
important implications for star formation in the inner bulge, since
a sphere (the inner bulge in this case) with a volume density profile
close to $r^{-1}$ has negligible
tidal forces in it (the $r^{-1}$ density profile gives no tidal forces;
if the density drop is shallower than $r^{-1}$, the tidal force even
becomes compressive).  Thus if the local volume density in the inner
bulge can be indeed shallower than, say, $r^{-1.5}$, formation of stars there
may not be as difficult as has been speculated (the galactic tidal field
acts against the collapse or contraction of molecular clouds into a star).
The enclosed mass
as a function of galactocentric radius deduced by \markcite{LHW92}Lindqvist
et al. (1992a; Fig.~10) shows a steep increase (thus shallow decrease
in density) between 30 and 100~pc (steeper than $\sim r^{1.5}$, implying
a density profile shallower than $r^{-1.5}$), and this supports
our simulation results that the density profile becomes shallower in the
$30 \lsim r/{\rm pc} \lsim 100$ region and also supports the above discussion
on the possibly favorable star formation environment there.

Considering the age of the Galaxy, OH/IR stars in the inner bulge
are relatively young.  Our comparison with a ``young'' population
strongly supports our hypothesis that the GMCs are responsible for the vertical
diffusion, but does not necessarily imply that the entire $r^{-2}$ stellar
population is a result of sustained star formation in the CMZ.
To address this issue, one needs to account for the evolution and
accumulation of stellar populations over the lifetime of the
Galaxy.  Since the evolution of aspect ratios in Figure~\ref{fig:allaspect}
appears to converge to an asymptotic value at times $\gsim 2$~Gyr,
we may expect that the agglomeration of stars formed in the CMZ
over the lifetime of the Galaxy will have a stellar distribution
not much different from that of the last stages of our simulations.
A systematic observation that covers the full inner bulge region
with good resolution and sensitivity is needed to take the next step
in this analysis.  This need may be fulfilled by the currently ongoing
2MASS survey.

%%%%%%%%%%%%%%%%%%%%%%%%%%%%%%%%%%%%%%%%%%%%%%%%%%%%%%%%%%%%%%%%%%%%%%%%%%%%%%%%
\section{SUMMARY}
\label{sec:summary}

We have performed numerical simulations of scattering of young stars
off the GMCs in the inner bulge.  Several different bulge potentials,
GMC models and distributions, and initial conditions of the stellar
population have been considered.

First we find that when $\alpha_{GMC} = -1$, the stars inside $\sim 70$~pc
experience rapid radial diffusion in the plane during the early phase.  The
consequences of this are that 1) later evolution of the inner stars is very
similar to that of the outer stars (initial $R \gsim 80$), and 2) the depletion
of stars in the inner region makes the volume density profile in the plane
between 30 and 100~pc quite shallow (possibly close to $r^{-1}$).  The latter
is supported by the rapid increase in the enclosed mass between 30 and 100~pc
(\markcite{LHW92}Lindqvist et al. 1992a), and implies that the tidal forces
in that region may not be as hostile to star formation as has been previously
conjectured.

After a rapid initial rise
during the first $\sim 0.5$~Gyr, especially for the inner stars,
$\bar R$ linearly increases afterwards (in case of model~1,
$\bar R$ reaches 150~pc in $\sim 3$~Gyr).  The vertical positional dispersion
of stars, on the other hand, grows almost linearly for the whole computational
interval (model~1 reaches $\sigma_z = 50$~pc at $\sim 3$~Gyr).

The projected surface density of test stars along the galactic plane
inside 100~pc decreases
as stars diffuse out both radially and vertically, but roughly maintains
its initial slope over our simulation period.  On the other hand, the slope
of the projected surface density profile along the $b$ (vertical) axis
becomes significantly shallower during the same period and makes
the overall stellar population rounder.

The comparison between observations and our simulations 
of the $l$-axis surface density profile (projected along the
line-of-sight) in the $10 \lsim r/{\rm pc} \lsim 100$ region suggests that 
newborn stars have an initial surface density profile (projected along the
$z$-axis in the plane) of $R^{-2} \,$--$\, R^{-1}$.
However, the estimation of 3-dimensional
structure from 2-dimensional observation requires more
careful study in the context of non-axisymmetric stellar models in order to
reveal the true volume density profile in the inner bulge.

The aspect ratios of our young stellar configuration are initially larger than
6, but the ratios become smaller than 3 in 1--$2$~Gyr, except for one model.
This is because the scattering off GMCs
diffuses stars preferentially in the vertical direction (since the density
gradient is greater in that direction), except in the very early phase.
The aspect ratios converge to values between 1.5 and 2.5 in 1--$2$~Gyr.
Since OH/IR stars in the inner bulge , which are estimated to be 1--2~Gyr old
(\markcite{Se99}Sjouwerman et al. 1999), have an aspect ratio of 2.6,
we conclude that scattering by the GMCs is one of the predominant mechanisms,
and possibly the most important for the vertical diffusion of stars in the
inner Galactic bulge.

\acknowledgements
S.S.K. is grateful to Jongsoo Kim and Kwang-Il Seon for help with
computer facilities, and to Burkhard Fuchs, Cheongho Han, Chang Won Lee,
Hyung Mok Lee, Douglas Lin, David Merritt, William Newman, and Roland Wielen
for valuable discussions.  We also appreciate the comments of an anonymous
referee, which significantly improved our manuscript.
This work was supported in part by the International Cooperative Research
Program of the Korea Research Foundation to Seoul National University in 1999,
and in part by a NASA grant to UCLA.

%%%%%%%%%%%%%%%%%%%%%%%%%%%%%%%%%%%%%%%%%%%%%%%%%%%%%%%%%%%%%%%%%%%%%%%%%%%%%%%%

\end{document}